\documentclass{article}

\usepackage{PRIMEarxiv}

\usepackage[utf8]{inputenc} % allow utf-8 input
\usepackage[T1]{fontenc}    % use 8-bit T1 fonts
\usepackage{hyperref}       % hyperlinks
\usepackage{url}            % simple URL typesetting
\usepackage{booktabs}       % professional-quality tables
\usepackage{amsfonts}       % blackboard math symbols
\usepackage{nicefrac}       % compact symbols for 1/2, etc.
\usepackage{microtype}      % microtypography
\usepackage{lipsum}
\usepackage{fancyhdr}       % header
\usepackage{graphicx}       % graphics
\graphicspath{{media/}}     % organize your images and other figures under media/ folder

\title{Comparative Study of Simulators for Vehicular Networks
}

\author{
  Rida Saghir \\
  University of Bremen \\
  Bremen, Germany\\
  \texttt{rida1@uni-bremen.de} \\
   \And
 Thenuka Karunathilake \\
  University of Bremen \\
  Bremen, Germany\\
  \texttt{thenuka@comnets.uni-bremen.de} \\
   \And
  Anna F\"orster \\
  University of Bremen \\
  Bremen, Germany\\
  \texttt{anna.foerster@uni-bremen.de} \\
}

\begin{document}
\maketitle

\begin{abstract}
Vehicular Adhoc networks (VANETs) are composed of vehicles connected with wireless links to exchange data. VANETs have become the backbone of the Intelligent Transportation Systems (ITS) in smart cities and enable many essential services like roadside safety, traffic management, platooning, etc with vehicle-to-vehicle (V2V) and vehicle-to-infrastructure (V2I) communications. In any form of research testing and evaluation plays a crucial role. However, in VANETs, real-world experiments require high investment, and heavy resources and can cause many practical difficulties. Therefore, simulations have become critical and the primary way of evaluating VANETs' applications.
Furthermore, the upfront challenge is the realistic capture of the networking mechanism of VANETs, which varies from situation to situation. Several factors may contribute to the successful achievement of a random realistic networking behavior. However, the biggest dependency is a powerful tool for the implementation, which could probably take into account all the configuration parameters, loss factors, mobility schemes, and other key features of a VANET, yet give out practical performance metrics with a good trade-off between investment of resources and the results. Hence, the aim of this research is to evaluate some simulators in the scope of VANETs with respect to resource utilization, packet delivery, and computational time. 
\end{abstract}

% keywords can be removed
\keywords{Vehicular Networks \and OMNeT++ \and ns-3}

\section{Introduction}
  \label{sec:intro}
  
VANET is a subset of Mobile Ad hoc networks (MANETs) which is formed by road vehicles and road infrastructure with spontaneous wireless connections to exchange any required information. VANETs have several distinguishable characteristics compared to MANETs, such as vehicular nodes having arbitrary movement patterns with no constant speed or direction with various possible topologies of the roads. The information relays among vehicles, and road infrastructure can cater to many vehicular application areas, namely, road safety, route management, navigation, and platooning. However, such systems are difficult to test in the real world or can bring in many challenges and life hazards and require many financial resources. Therefore, simulating vehicular behavior in a simulation tool is beneficial in many ways and still can obtain realistic results because of the many available real-world models inside the simulator. This is achieved with the combination of different tools, including simulators, mobility models, favorable routing protocols, real-life traces, suitable communication technology, and lots of other factors making up a whole framework. \par

The scope of VANETs simulations covers two main phases:
\begin{itemize}
  \item Mobility Simulations.
  \item Network Simulations.
\end{itemize}

In the scope of vehicular networks, mobility simulation fabricates suitable movement patterns of the nodes (vehicles) based on the virtual road boundaries or lanes, depicting the behavior of actual vehicles in real life. Some extra features like traffic lights and speed breakers can also be introduced depending on the scalability of the simulator. On the other hand, the network simulation determines the behavior of data transmission depending on the selected network scenario, communication technology, and mobility model and includes all other key parameters that make up a network topology like; transmit power, message size and type, sending intervals etc. 

This research aims to find out and evaluate available simulators in the scope of VANETs that are also readily available and can deliver realistic results. For this purpose, a survey of different research papers and projects was conducted to find the available simulators for VANETs. We have identified more than 10 simulators that can be used for our analysis. However, due to many practical constraints, which will be discussed in later sections, finally, we decided to use SUMO as the mobility simulator, and ns-3 and OMNeT++ as the network simulators.

The rest of the paper is organized as follows. In Section \ref{sec:related} we discuss the research on different simulators and how often they were cited in publications. Section \ref{sec:chosensimulators} provides details about the selected simulators in our work. Section \ref{sec:eval} presents our performance evaluation on the selected simulators, while Section~\ref{sec:shortcoming} discusses our first-hand user experience on the simulators. Finally, Section~\ref{sec:conc} concludes the paper.

\section{Related Work}
\label{sec:related}

\begin{figure}[htp!]
    \centering
    \includegraphics[width=0.6\textwidth]{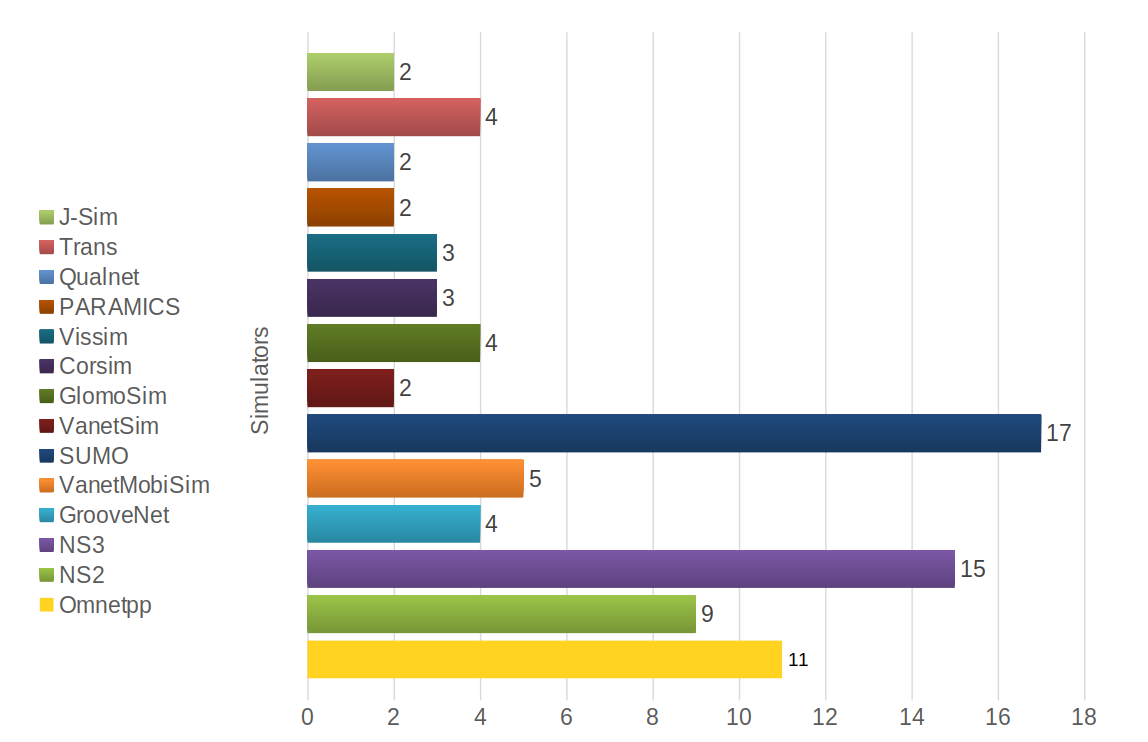}
    \caption{Available VANET Simulators with number of citations}
    \label{fig:mentionedsimulators}
\end{figure}

The demand for VANET applications has been rising for the past two decades. Therefore multiple simulators have been developed, worked, and researched. In our analysis, we reviewed around 80 scientific publications to find out about the available simulators. Some of them are \cite{9070933}, \cite{4781775}, \cite{6903733}, \cite{8958692}, \cite{harri2006vanetmobisim}, \cite{barnett1992netsim}, \cite{veith1999netsim}, \cite{lan2011open}, \cite{amoozadeh2019ventos}, \cite{nayak2014comparison} and \cite{nayak2015analysis}. In \cite{9070933}, the authors discuss various network simulators and mobility generators for VANETs, along with a precise comparison between them based on their features and working behaviors. The authors of \cite{kabir2014detail} have also compared numerous network simulators like ns2, ns3, QualNet, and OMNET++ in terms of their basic features and the advantages and limitations that they all offer.
Furthermore, in \cite{minakov2016comparative}, the authors describe different simulation tools and their run-time performances in the scope of wireless sensor networks. Figure \ref{fig:mentionedsimulators} depicts the number of citations of each simulator we identified. These simulators either work as mobility simulators or network simulators. However, some of the mentioned simulators could handle both mobility and network abilities; therefore, they are regarded as integrated simulators.

NETSIM \cite{veith1999netsim} \cite{barnett1992netsim} is an end-to-end, full-stack, packet-level network simulator and emulator developed by TETCOS. It is used to model discrete event network simulations using the event graph approach and is entirely written in JAVA. Groovenet is a street-based hybrid simulator that allows communication between simulated vehicles, real vehicles, and between simulated and real vehicles such that they can exchange data as mentioned in \cite{4141794}. Glomosim \cite{bajaj1999glomosim}, is an open-source, library-based, discrete event simulator that is used for network protocol simulations based on wired and wireless network systems. VanetMobiSim \cite{harri2006vanetmobisim}, a type of mobility simulator and an extension of CanuMobiSim which itself is a framework for user mobility modeling. Qualnet \cite{dinesh2014qualnet}, a commercial network simulation software that is used to create network digital twins.    

As shown in Figure \ref{fig:mentionedsimulators}, there are various simulators to work with but the fact that many of these are not under active development or do not have very good technical support in terms of documentation or tutorials, appear as a big barrier to use them. A few examples are J-Sim, Trans, VanetMobiSim, VanetSim, etc. We decided to analyze the most cited simulators with readily available supporting materials. Therefore, we selected to analyze and compare OMNET++ (using Veins framework), and ns-3 as network simulators along with utilizing SUMO as a mobility simulator as they were considered to be suitable in the context of vehicular networks after going through \cite{hassan2009vanet}, \cite{weber2021vanet}, \cite{7284922} and \cite{sommer2015simulation}.

Our work is significantly different from the work mentioned above because we solely focus on VANETs and explore the simulators' scalability, which is a significant problem statement when simulating large and dense VANETs. Furthermore, we thoroughly analyze the required resources by the different simulators and their run-time performance while tackling large-scale networks. We believe our findings will be beneficial for the research community in the future to select the optimal simulator for VANET simulations.

\section{Selected Simulators}
\label{sec:chosensimulators}

In our experiments, as mentioned in the \ref{sec:related} we selected SUMO, ns-3, and OMNeT++. They were chosen because of the abundance of developer support, usability in the context of vehicular networks, available tutorials, proper active QA forums, and active development in terms of bugs and versions. Also, open-source simulators were prioritized. 

\subsection{SUMO}
In order to have realistic VANET results, it is necessary to have a realistic vehicle movement pattern as in urban areas, heavy distribution of vehicles with specific traffic management behavior so the simulations could depict close to a real environment. SUMO (Simulation of Urban MObility) \cite{8239038} \cite{8569938} is an open-source microscopic traffic simulator that is used to generate vehicular movement patterns under a specific trace by creating manually or by importing road networks from OpenStreetMap, VISUM etc. It is now heavily engaged in simulating VANETs because of its ability to support traffic management, multimodal traffic, traffic lights, automated driving, vehicle communication, etc. It also carries the microscopic label, which means that each vehicle and its dynamics are modeled individually, making SUMO an even distinct choice among other mobility simulators.

\subsection{OMNeT++}
\label{sec:OMNET}
OMNeT++ (Objective Modular Network Testbed in C++) is an object-oriented, open-source, discrete event simulator that has been available publicly since 1997 for private and academic use. It is a component-based C++ simulation library that has gained widespread popularity for building network simulators. It is a modular framework which means that the small components are programmed in C++ and combined into larger components using a high-level language NED (Network Description). This modular property enables the re-usability of models. Along with the extensive GUI support, the embedding of the simulation kernels into the applications is among the useful features. OMNeT++ was developed in a generalized manner so that it could cover various application areas like wired and wireless networks, on-chip networks, queuing networks, multiprocessors, and other distributed or parallel systems.\par
Although OMNeT++ is developed with C++, it is still possible to develop your own modules in other languages like Java and C and test them in simulations in collaboration with built-in modules to check their authenticity. Various individual frameworks and models, like the INET framework and Mobility framework, are being developed and maintained independently and can support specific application areas. Different developers and research groups work on the frameworks to serve areas like wireless and ad-hoc networks, sensor networks, IP networks, etc. An overview of OMNeT++ can be seen in \cite{varga2008overview} and \cite{inproceedings}.

\subsubsection{OMNeT++ as a Vehicular Network Simulator}
As discussed earlier, there are various individually developed frameworks available that serve application-specific areas. OMNeT++ is not a network simulator but only a discrete event simulator but plugging in a framework provides OMNeT++ with the modules sufficient to simulate different network models. That framework would be INET which provides help with protocols, agents, and other supporting modules for internet stack, wired, wireless link layer, mobility models, and several other applications. Different developers tend to contribute to this framework over time to make it capable of a wide range of networks. Now, a framework named Veins can be used to make the test bed even better suited for VANETs. The behavior of vehicular nodes and their mobility or packet exchange differs from that of MANETs. Therefore, Veins, an open-source simulation framework, integrates the VANET mechanism in coordination with OMNeT++ and SUMO. An overview of Veins can be found in \cite{sommer2019veins}. 

\subsection{ns-3}
\label{sec:ns-3}
ns-3 is also a discrete event simulator for Internet systems, wireless and wired networks, and much more. It is a sequel of ns-1 and ns-2 and was created in 2008. It is built with C++ on cores and has Python bindings. The simulation workflow includes; creating the network topology, models, nodes, and link configuration, executing the simulation, performance analysis, and graphical visualization. ns-3 is an improved version of ns-2 with respect to realism in software implementation. ns-3 is developed similar to the architecture of a Linux System, with internal interfaces (network to device driver) and application interfaces (sockets) that map well to how computers are built today. There is a significant amount of materials available on the official website, and also several papers have been written \cite{9121007}, \cite{6103008}, and \cite{7284922}. 

\subsubsection{ns-3 as a Vehicular Network Simulator}
As ns-3 is a network simulator, it does not require additional libraries like INET. Also, a vehicular simulation example is available that can implement routing protocols suitable for VANETs like OLSR, AODV, DSDV, and DSR and have predefined map samples in it. The example simulation checks for the vehicular nodes in the communication range and the same channel to be able to exchange packets, depending on the selected routing protocols. The example has most of the parameters available to be altered. For example, Propagation Loss, routing protocols, node speed, etc. Also, ns-3 can be coupled with SUMO like OMNET++, and the sample map can be replaced by the one made by the user. However, the coupling can be online as well as offline.

\begin{itemize}
    \item \textbf{Offline Technique}: The trace files can be generated from SUMO that captures the mobility and then can be converted to a trace file that is readable by ns-3 with the help of NS2MobilityHelper.
    \item \textbf{Online Technique}: Using tools like iTETRIS or VSimRTI that combine SUMO and ns-3 live. However, not much technical support is available for this technique. Also, the main developers of ns-3 prefer to do it with the offline technique. 

\end{itemize}

\section{Experiments and Performance Analysis}
\label{sec:eval}

The goal of the research is to compare the performances of both simulators in the scope of VANETs. For a precise comparison and evaluation, a typical simulation scenario is used, which is fed by realistic vehicle traffic captured in the suburban map of Bremen City, Germany. Table \ref{table:simulation parameters} shows the simulation parameters that were considered in our experiments, after reviewing different papers which include VANETs simulations \cite{8320746}, \cite{7149649}, \cite{8596531}, \cite{9234202}, \cite{parmar2014performance}, \cite{6103008}, \cite{7154825}, \cite{pandeyimprovisation} and \cite{bilalb2013performance}.

\begin{table}[h!]
\centering
\begin{tabular}{ ||c c|| }
 \hline
 \multicolumn{2}{|c|}{Simulation Parameters} \\
 \hline
 Map   & Bremen Horn - University of Bremen \\
 Edge Length&   41.05 km\\
 Lane Length & 43.25 km\\
 Transmit Power    &20dBm\\
 Routing Protocol&   AODV \\
 Vehicle Speed& 14 m/s\\
 Routing Rate& 2.048kbps\\
 Number of Nodes& 50-600 (around)\\
 Simulation Time& 50-400 sec\\
 Mobility Generator& SUMO\\
 Loss Model& Free space Loss Model\\
 Propagation Delay Model& Constant Speed Propagation Delay\\
 
 \hline
 \end{tabular}
\caption{Simulation Parameters}
\label{table:simulation parameters}
\end{table}

\begin{figure}[htp]
    \centering
    \includegraphics[width=0.6\textwidth]{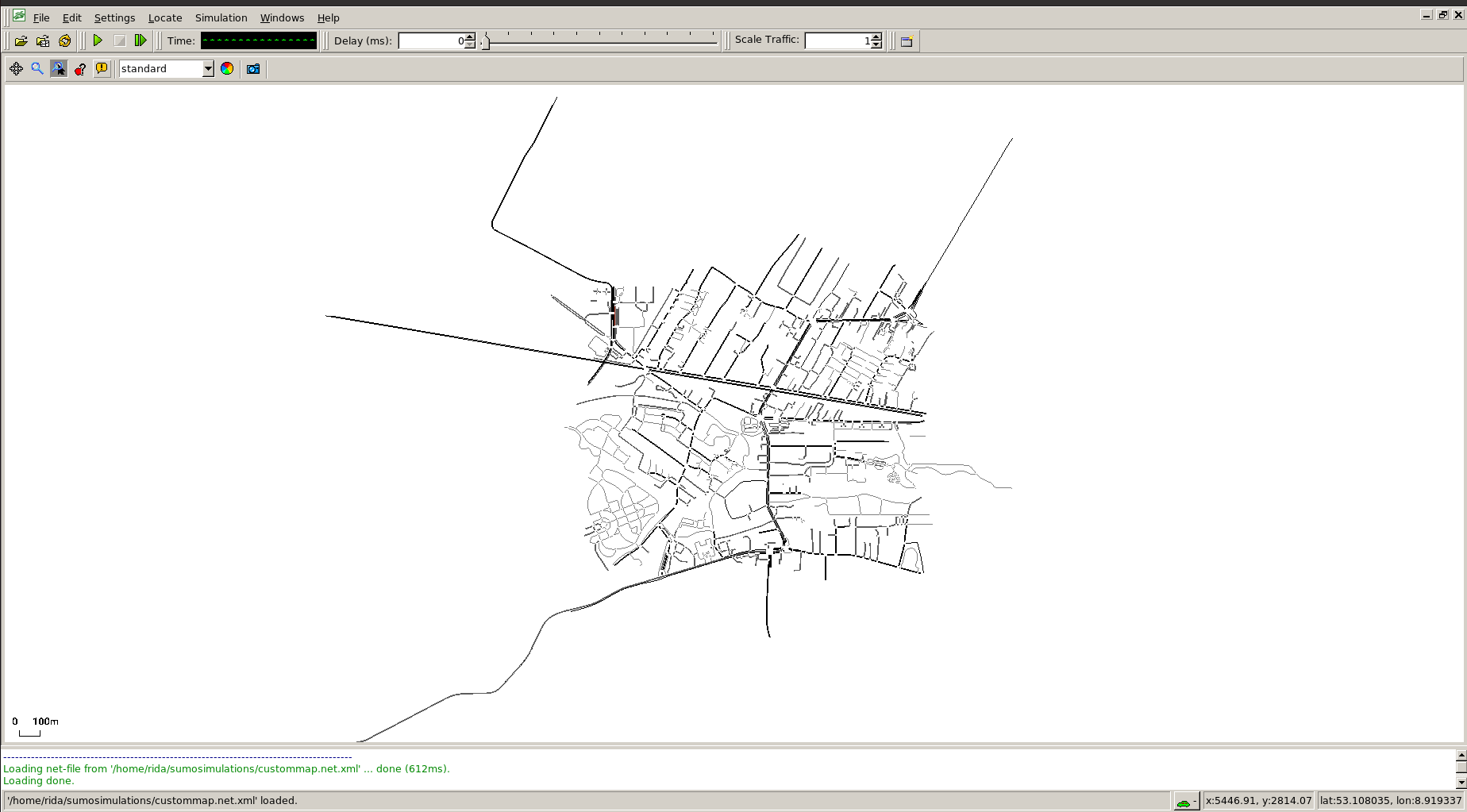}
    \caption{SUMO Network used for the simulations}
    \label{Abb:sumonetwork}
\end{figure}

The simulations were run 3 times for each setting, and the average value was taken for the comparison. Figure \ref{Abb:sumonetwork} shows the dense network architecture imported from Google Maps that is used for this experiment. The vehicles were added to the network randomly at different times. Therefore, the number of nodes in the networks depends on the simulation time. We simulated the network for 50, 100, 200, 300, and 400 seconds, and in each of these simulations times, there were different numbers of nodes in the network. Therefore, the number of nodes varies approximately between 50 for the 50-second simulation and 600 for the 400 second simulations. Due to the long computational time of the simulators, we limit our simulations to 400 seconds. We compared the simulators in terms of memory usage, computational time, and packet delivery ratio. 

\subsection{Memory Usage}
\label{sec:Memory usage}

\begin{figure}[htp]
    \centering
    \includegraphics[width=0.6\textwidth]{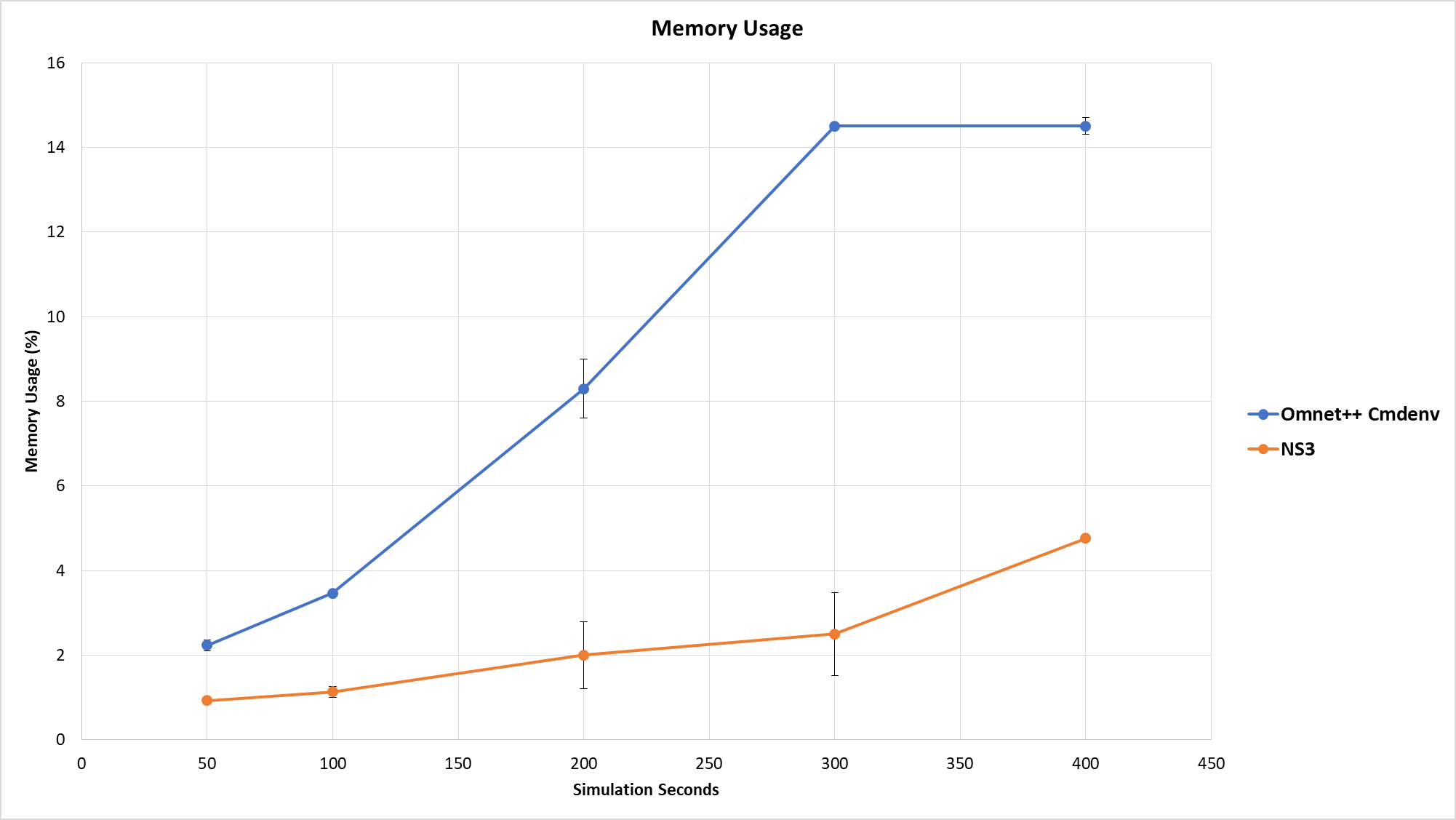}
    \caption{The amount of memory usage (\%) vs the simulation duration (seconds)}
    \label{Abb:memoryusage}
\end{figure}

The memory usage was recorded through a Python script which recorded values every second throughout the simulation run. Both simulators were used through command line terminals, which contain the least overhead. This can be useful when one has to simulate quite large networks without compromising too many resources. ns-3 outperforms OMNeT++ in terms of memory usage as shown in Figure \ref{Abb:memoryusage}. The difference between the two simulators tends to increase with increasing simulation seconds. The reason for the results can also be the minimal tracing ability of ns-3 which doesn't require the user to generate vector and scalar files with graphs, node-to-node performance, and histograms for basic network performance evaluation, enabling it to generate some top-shelf results without dealing with detailed result hierarchy of nodes or different communication layers. However, the memory usage in OMNet++ can be reduced if the results recording can be minimized in OMNeT++. 

\subsection{Computational Time}
\label{sec:Computational Time}

\begin{figure}[htp]
    \centering
    \includegraphics[width=0.6\textwidth]{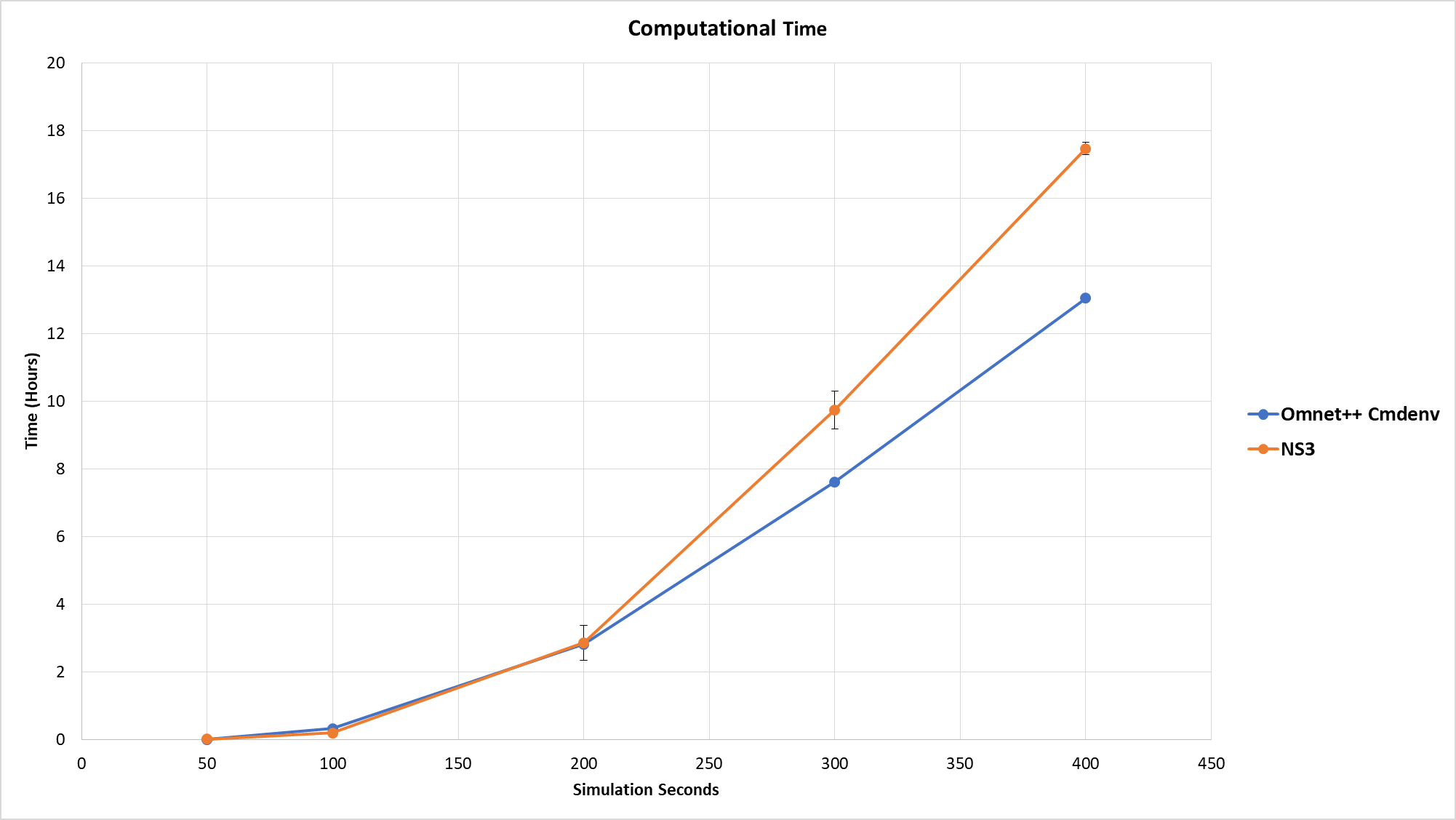}
    \caption{The computational time (hours) vs the simulation duration (seconds)}
    \label{Abb:computationaltime}
\end{figure}

The time required by each simulator to complete the simulation was considered as the computational time, and the obtained results are depicted in Figure \ref{Abb:computationaltime}. In the simulations shorter than 200 seconds, both simulators required approximately the same number of hours to complete the simulation run. However, for longer simulations, the results were significantly different for the two simulators. Furthermore, the difference tends to increase with the increasing simulation time. As can be observed, OMNeT++ takes less than 8 hours for 300 seconds, whereas ns-3 takes around 9.5 hours. OMNeT++ could simulate 400 seconds in around 13 hours, and ns-3 took around 17.5 hours. The reason for this behavior could be the structure, implementation, and computational complexity of the ns-3 wireless module for 802.11, which makes ns3 invest more time per simulation step. This issue has been discussed in detail in \cite{6240251} with some possible solutions. 

\subsection{Packet Delivery Ratio}
\label{sec:Packet delivery ratio}

The packet delivery ratio decreases with increasing simulation time in both simulators.
The packet delivery ratio was around  76\% in ns-3 and 87\% in OMNeT++ for the 50 seconds simulation. However, during the simulation of 100 seconds, the packet delivery ratio dropped significantly to 4.2\% in ns-3. Similarly, the delivery ratio dropped for OMNeT++ but not as much as in ns-3 and remained around 41\%. For the following simulations also, the delivery ratio decreased gradually for both the simulators. However, in all of the scenarios, OMNeT++ manages to maintain a better delivery ratio than ns-3. The reasons could be the slow processing and decision-making of the possible receptions, due to which the packets are probably getting wasted. Another reason could be the population of nodes from the start in ns-3. In the case of OMNeT++, the vehicles are gradually being added to the simulation; therefore, the throughput is gradually affected by the elapsed time. However, this is not the case with ns-3 as it follows the trace file and all the possible nodes are created at the start of the simulation (through the configuration file) and act as wireless nodes from the beginning, but they start moving at the configured time by SUMO.

\begin{figure}[htp]
    \centering
    \includegraphics[width=0.6\textwidth]{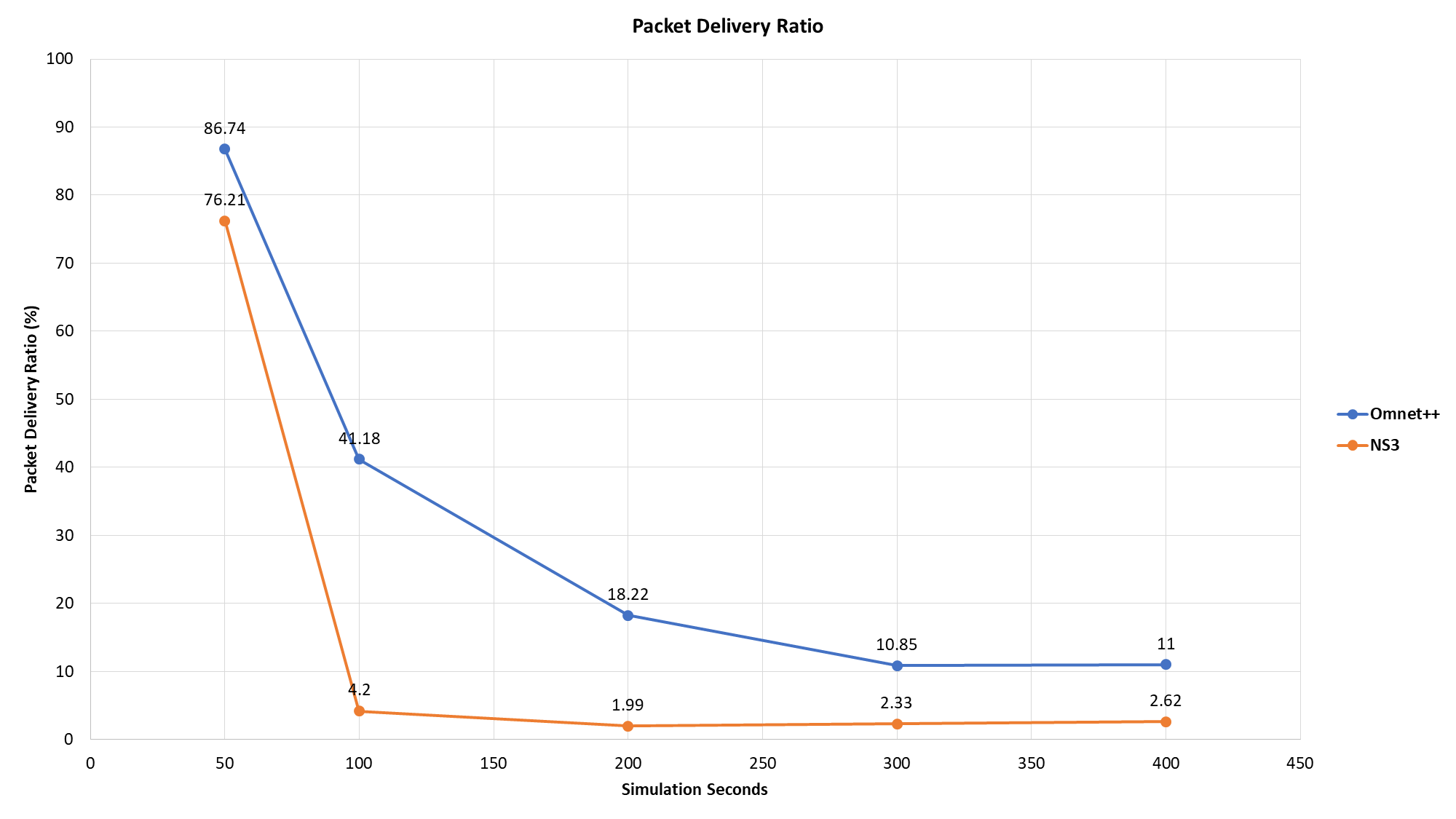}
    \caption{Packet delivery ratio (\%) vs the simulation duration (seconds)}
    \label{Abb:packetdeliveryratio}
\end{figure}

\section{Shortcomings Of The Simulators}
\label{sec:shortcoming}

In this section, we describe our experience with the two network simulators. We discuss the problems that were identified during the experiments. This information will be helpful for the research community when selecting a suitable simulator. Furthermore, in Table \ref{table:Comparison table}, we have summarized and compared the main features of OMNeT++ and ns-3. The learning level mentioned in Table \ref{table:Comparison table} refers to the fact that deploying a network scenario in ns-3 requires very good programming knowledge.
On the contrary, in OMNeT++, this process is made easier with the help of NED and .ini files. However, managing the workspace is complicated and is prone to corruption in OMNeT++. Data collection (especially node-level information) in ns-3 is not very efficient and often requires external tools like trace metrics, Wireshark etc. Sometimes, some lines of code might be needed for the result acquisition. However, in OMNeT++, the results can be easily acquired filtered, and categorized as histograms, vectors, and scalars. Also, OMNeT++ is only a discrete event simulator which requires additional plugins to work as a network simulator. However, this is not the case with ns-3, and it consists of all the required network elements.

\begin{table*}[h!]
\begin{center}
\centering
\footnotesize
\begin{tabular}{ |c|c|c| }
  \hline
 \textbf{Features}& \textbf{ns-3} & \textbf{OMNET++} \\ \hline
 Language Supported& C++,Python Bindings& C++,NED (High level language) \\ \hline
 Operating System& GNU/Linux, MacOS& Windows, Linux, MAC OS/X \\ \hline
 GUI Support& Poor (Operation from terminal)& Good (Network Animator/Project Explorer\\ \hline
 Last Updated& July 2023& September 2022\\ \hline
 License& Open source & Open source and Commercial \\ \hline
 Learning Level& Normal & Easy \\ \hline
 Propagation Models& 43& 10 \\ \hline
 Wireless Access for Vehicular Networks& IEEE80211p, IEEE1609.4& IEEE80211p, IEEE1609.4\\ \hline
 Technical Documentation& Available& Available and detailed\\ \hline
Dependency to interact with SUMO& VSimRTI, iTETRIS& Veins\\ \hline
Data Collection& Complicated (External tools required)& Easy (Built-in result filters)\\ \hline
Routing Protocols& OLSR, AODV, DSDV, DSR& AODV, DSDV, OLSR, GPSR\\ \hline
Mobility Models& 11& 14 \\ \hline
Support for internet stack and other protocols& Built in& INET \\ \hline
 \end{tabular}
 \end{center}
\caption{Comparison Table}
\label{table:Comparison table}
\end{table*}

\subsection{Problems with OMNET++}

\begin{itemize}
    \item Routing protocols are not readily implemented for VANETs when using the Veins framework. Some modifications are required that can lead to unidentified errors. Furthermore, some routing protocols suitable for VANETs are missing in the INET framework, which leaves a lot of work for programmers as the routing protocols are mainly focused on MANET (Mobile Adhoc Network). 
    \item Version conflicts between OMNeT++, INET, and Veins have to be investigated. Since to make OMNeT++ work as a vehicular and network simulator, one has to plugin two frameworks, Veins and INET. However, version compatibility of all three simultaneously is a complicated task, especially when your program requires something specifically from a different version of any framework.
    \item The workspace is prone to get corrupted over time. OMNeT++ uses the Eclipse platform, which is known to sometimes cause unidentified problems with the workspace of projects. Therefore, one has to be careful about their selected locations and backups. This increases the required efforts for maintaining a project. 
    \item Veins framework compromises some functionality when working with INET framework. The simulation environment or INET does not know if any nodes are created or destroyed during the simulation by Veins as per the used mobility patterns. Hence, this causes problems when those nodes are selected as exclusive destinations.  
\end{itemize}

\subsection{Problems with ns-3}
\begin{itemize}
    \item As ns-3 does not come along with a project workspace, hence working and tracing back supporting modules and libraries causing network bugs is difficult. Even if somebody used debugging, finding the relevant files and lines of code is a very tedious task.
    \item Cannot visualize run time issues without going into debug mode. Also, the error codes are vaguely explained in the error message. 
    \item Very few active maintainers who can respond to the queries. 
    \item Modifications according to vehicular applications are difficult as few repositories or examples are available online. It has only one VANET example, which is programmed to send mandatory safety messages. However, lots of programming knowledge is required to make the application act differently.  
    \item ns-3 tends to consume much more time per simulation step for VANET simulations, as also mentioned in \cite{6240251}
    \item Analyzing node-level information takes so much time and are saved in the form of PCAP files per node. Therefore, consumes a lot of space on the disk.
    \item The programs are very lengthy and not easily customizable or readable. Unlike OMNeT++, which has separate NED and .ini files for configurations.  
\end{itemize}

\section{Summary and Conclusions}\label{sec:conc}

In this paper, we searched for available VANET simulators. ns-3 and OMNeT++ were identified as the most common network simulators coupled with the mobility simulator SUMO. We compared and analyzed the performance of ns-3 and OMNeT++ in VANET simulations involved in a general city map with typical heavy vehicle flow with AODV routing protocol. From the simulation results, we gained a perception of the differences between the simulators, especially when it comes to the simulation of VANETs. We analyzed their computational time, memory usage, and packet delivery ratio to compare their resource consumption behavior and performances. Various other pros and cons were also analyzed during the process. It was also noticed that the results could be different even by keeping the simulation parameters identical for both simulators.

%Bibliography
\bibliographystyle{unsrt}  
\bibliography{templateArxiv}

\end{document}